\documentclass[
    ,final            
  ]
  {aipproc}

\layoutstyle{6x9}
\def\met{\mbox{$\rm {\hbox{E\kern-0.4em\lower-.1ex\hbox{/}}}_T$}}
\def\mpt{\mbox{$\rm {\hbox{P\kern-0.4em\lower-.1ex\hbox{/}}}_T$}}
\def\metx{\mbox{$\rm {\hbox{E\kern-0.4em\lower-.1ex\hbox{/}}}_x$}}
\def\mety{\mbox{$\rm {\hbox{E\kern-0.4em\lower-.1ex\hbox{/}}}_y$}}
\def\@\fnsymbol#1{\ensuremath{ifcase#1\or \dagger\or
        \mathsection\or \ddagger \else @ctrerr\fi}}

\begin{document}

\title{SUSY Higgs Searches at D\O\ , Tevatron}


\author{Prolay Kumar Mal\\
(for D\O\  Collaboration)\\}{
  address={
  University of Notre Dame,\\
  Notre Dame, Indiana 46556, USA}
}

\begin{abstract}
During Run II of the Tevatron collider, D\O\  collaboration has
made extensive searches for the neutral MSSM Higgs bosons ($\phi$),
produced in $\rm p\bar{p}$ collisions at $\rm \sqrt{s}=1.96$ TeV.
Two such analyses, addressing inclusive $\phi$ production with
$\phi\rightarrow\tau^+\tau^-$, and associated $\phi b(\bar{b})$
production
with $\phi\rightarrow b\bar{b}$ are reported here.
No excess of events above the background expectation has been observed
in any of these analyses. The results are combined to set constraints
on the MSSM parameter space. 
\end{abstract}

\maketitle


\section{Introduction}
The Standard Model (SM) of particle physics
has been observed to be a consistent theory of
fundamental particles and their interactions up to
the energies they have been studied although 
one remaining essential ingredient the Higgs boson is yet to
be discovered.
Theoretical considerations teach us that SM
can not be the ultimate theory of elementary particles
and their interactions.
%
Amongst many other compelling theories,
the Minimal Supersymmetric extension of the SM (MSSM) is
one of the viable theoretical frameworks which
has the potential to overcome some of the
short comings of the SM by incorporating minimal supersymmetric
particle spectrum.
\par
Because of its supersymmetric structure, the MSSM\cite{mssmtheory2}\cite{mssmtheory}
requires at least
two Higgs doublets to generate masses to both ``up''-type and
``down''-type quarks (and the respective charged leptons). Such
a two-Higgs-doublet model predicts five physical Higgs bosons:
two CP-even Higgs bosons, h and H, one
CP-odd Higgs bosons, A and a pair of charged Higgs
bosons, $\rm H^\pm$. The MSSM Higgs sector at the lowest order
can fully be described in terms of one Higgs boson mass
($\rm M_A$ is chosen in CP-conserving scenario and $\rm M_{H^\pm}$
in CP-violating scenario) and $\rm tan\beta=v_2/v_1$,
where $\rm v_2(v_1)$ refers to the Higgs field that couples 
to the ``up''(``down'')-type quarks. However,
additional parameters {\it viz.,}
$\rm M_{SUSY},\  M_2,\   \mu\   and\    m_{\tilde{g}}$
enter at the level of radiative corrections.
It is to be noted that the coupling of the Higgs boson A
to ``down''-type quark such as bottom quark is proportional
to $\rm tan\beta$. So with respect to the SM, the production
cross section for A in association with bottom quark(s) gets
enhanced by a factor of $\rm tan^2\beta$. Furthermore, at
large $\rm tan\beta$ ($\rm \approx50$), there is mass degeneracy
between the Higgs bosons A and h or H depending on the value
of $\rm m_A$. Therefore, the total
production cross section for MSSM neutral Higgs bosons
{\it i.e.,} $\rm h/H/A\equiv\phi$ is twice that of the Higgs
boson A. Previously, LEP experiments have set the mass
of light Higgs boson $\rm m_h$ to be greater than
92.8 GeV\cite{LEPLIMIT} at 95\% CL.
\par
The MSSM neutral Higgs bosons ($ \phi$) mostly decay
into $ b\bar{b}$ pairs (90\%) or into $\tau^+\tau^-$
pairs ($\rm \sim 8\%$). So, at the Tevatron
$ p\bar{p}\rightarrow\phi(\rightarrow b\bar{b}) b(\bar{b})X$ 
and $ p\bar{p}\rightarrow\phi(\rightarrow\tau^+\tau^-) X$
processes are considered to be the most promising channels to
look for the signature of MSSM neutral Higgs bosons. Here we
report on the searches for MSSM neutral Higgs bosons
in the above channels performed by the D\O\ experiment
\cite{d0detector}. The combined results are
interpreted in different MSSM benchmark
scenarios\cite{mssminterpre}: ``$\rm m_h^{max}$'' and
``no-mixing''.

\section{Search for $\mathbf{p\bar{p}\rightarrow\phi (\rightarrow b\bar{b}) b(\bar{b})X}$}
The analysis focuses on MSSM neutral Higgs boson production
in association with one or two bottom quark(s) resulting in
three or four bottom quarks in the final state while,
CP-conservation in the Higgs sector is assumed.
260 $\rm pb^{-1}$ of D\O\  data collected between November 2002
and June 2004 have been utilized for this analysis. The data
are first filtered using a dedicated on line trigger designed to
maximize the signal acceptance. 
The Secondary Vertex (SV) tagging algorithm  has been used
to select the b-jets.
The D\O\  SV b-tagging algorithm selects
central b-jets ($\rm p_T>35$ GeV) with an efficiency of $\rm\approx55\%$
while the mis-tag rate for similar light quark jets is about 1\%.
\par
For Higgs masses between 90 and 150 GeV, the
$ hb\rightarrow b\bar{b}b$ signal events have been generated
using the Pythia~\cite{pythia} event generator followed by the full
D\O\  detector simulation and reconstruction. The Pythia generated
$\rm p_T$ and rapidity spectra of the Higgs bosons have been
adjusted to those from NLO calculations~\cite{NLOcal}. The
largest background contribution from 
QCD multijet processes have been determined directly
from data while contributions from other background
processes such as $\rm t\bar{t}$, $ Z^0(\rightarrow b\bar{b})$+jets
have been determined from Monte Carlo.
\par
\begin{figure}[h]
\begin{tabular}{cc}
\includegraphics[height=.2\textheight]{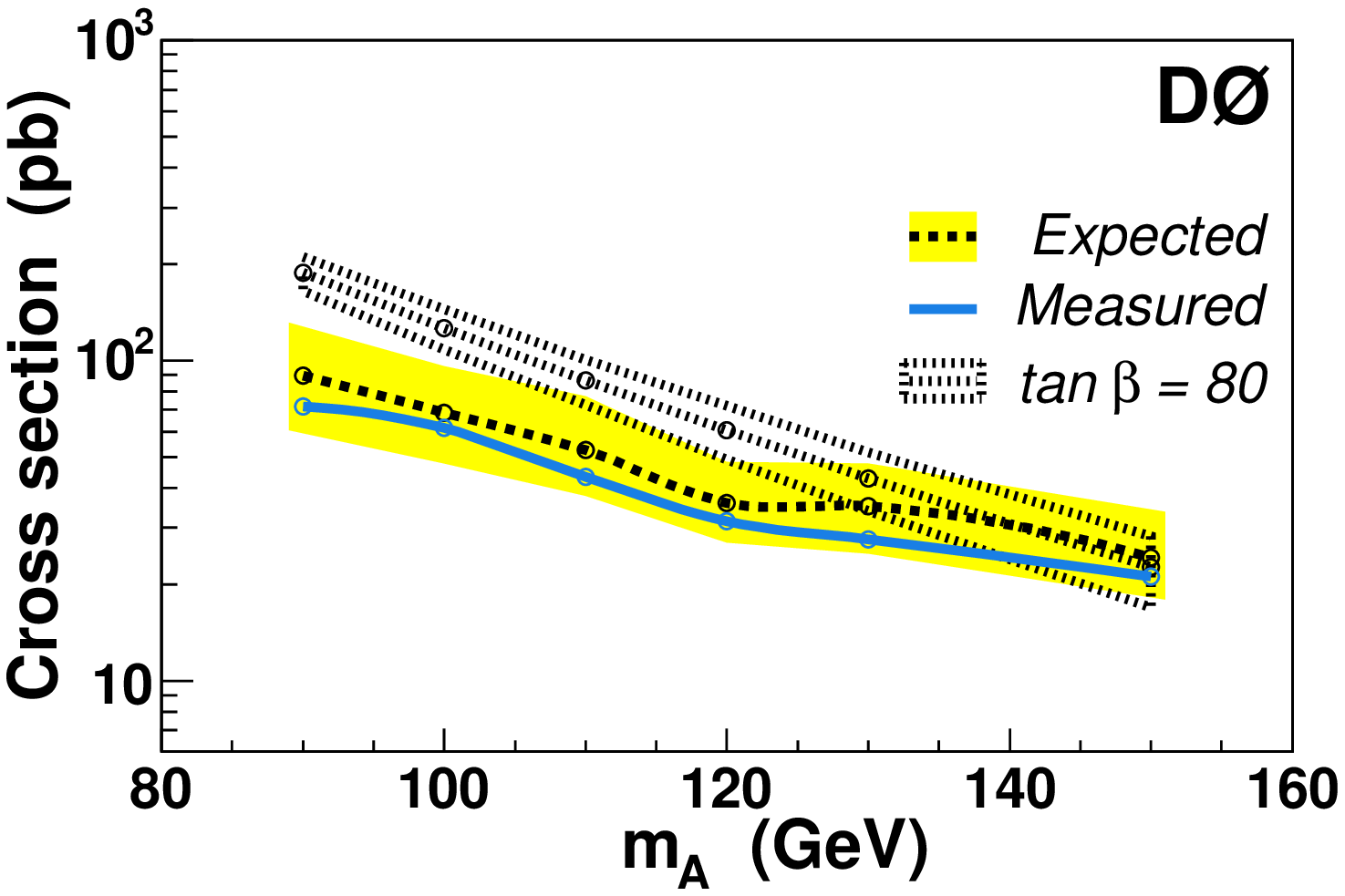}
&\includegraphics[height=.2\textheight]{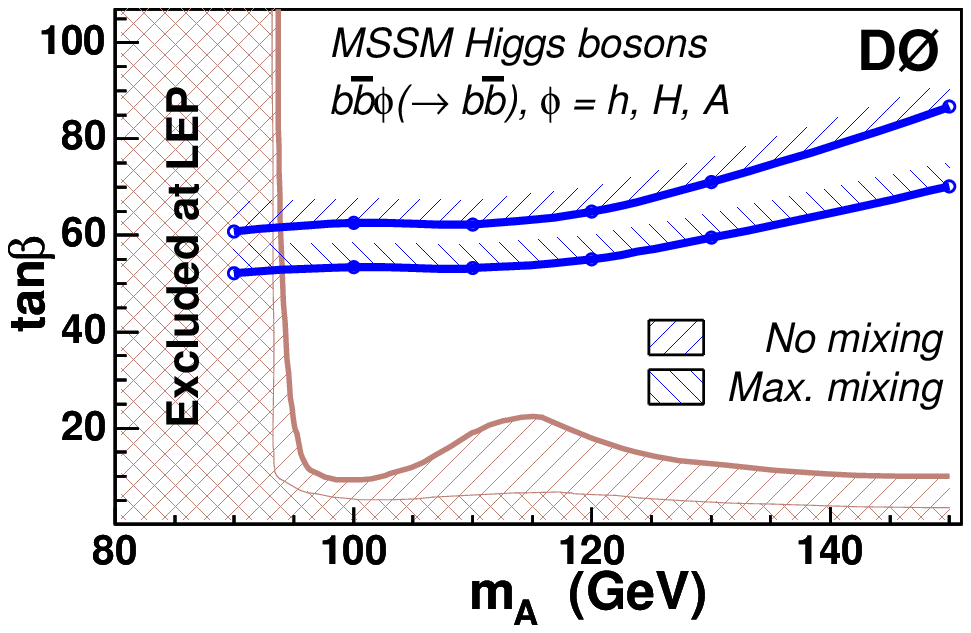}\\
(a) &(b)\\
\end{tabular}
\label{fig:hbbxsec}
\caption{(a) The upper limits on the production cross section as a
function of Higgs mass and (b) the limits on $\rm tan\beta-m_A$ plane
at 95\% CL.}
\end{figure}

The selected events are required
to have at least three b-tagged jets (within $\rm |\eta|<2.5$)
with $\rm p_T>$ 35, 20 and 15 GeV respectively for the
leading $\rm p_T$ jets.
The invariant mass distribution of the two leading
$\rm p_T$ jets in data is then compared with that
of the background expectation and no excess has been observed.
Fig.~\ref{fig:hbbxsec}(a) shows the limit on the production
cross section at 95\% CL, as a function of $\rm m_A$ for $\rm tan\beta=80$
in ``no mixing'' scenarios.
The results are also interpreted in the ``maximal mixing'' case and
the limits in $\rm tan\beta-m_A$ plane for both scenarios are
shown in Fig.~\ref{fig:hbbxsec}(b).
\section{Search for $\mathbf{p\bar{p}\rightarrow\phi (\rightarrow\tau^+\tau^-)X}$}
The analysis is based on 325 $\rm pb^{-1}$ of
data collected by the D\O\  experiment during the period September 2002
to August 2004. 
The signal consists of a pair of tau leptons. One of the tau leptons
is required to decay leptonically
into electron or muon, leading to three final state signature:
$ e\tau_h$, $ \mu\tau_h$ and $ e\mu$, where $ \tau_h$
represents a hadronically decayed tau lepton. The three types
of hadronically decayed $\tau$ leptons,
\begin{itemize}
\item $\tau$-type 1: a single track with a calorimeter cluster without any
electromagnetic sub clusters (1-prong, $ \pi$-like)
\item $\tau$-type 2: a single track associated with a calorimeter cluster and electromagnetic
sub clusters (1-prong, $ \rho$-like)
\item $\tau$-type 3: two or at least three tracks with invariant mass below 1.1 or 1.7 GeV
respectively (3-prong)
\end{itemize}
have been utilized in this analysis. A set of neural
networks\footnote[2]{Identical to the ones used for
$Z^0/\gamma^*\rightarrow\tau^+\tau^-$ cross section
measurements\cite{ztauxsec}}, one
for each type have been used for further discrimination over the
backgrounds.
\par
The signal as well as various background processes
have been
generated using Pythia. Events are then passed through the
full chain of D\O\  detector simulation and reconstruction
software. Apart from the QCD multijet background where a
jet mimics an electron/muon from tau lepton, all
other background processes are normalized to NLO and NNLO ($ Z^0$
boson, $ W^\pm$ boson, Drell-Yan and di-boson) cross sections.
\par
\begin{figure}[h]
\begin{tabular}{ccc}
\includegraphics[width=1.75in]{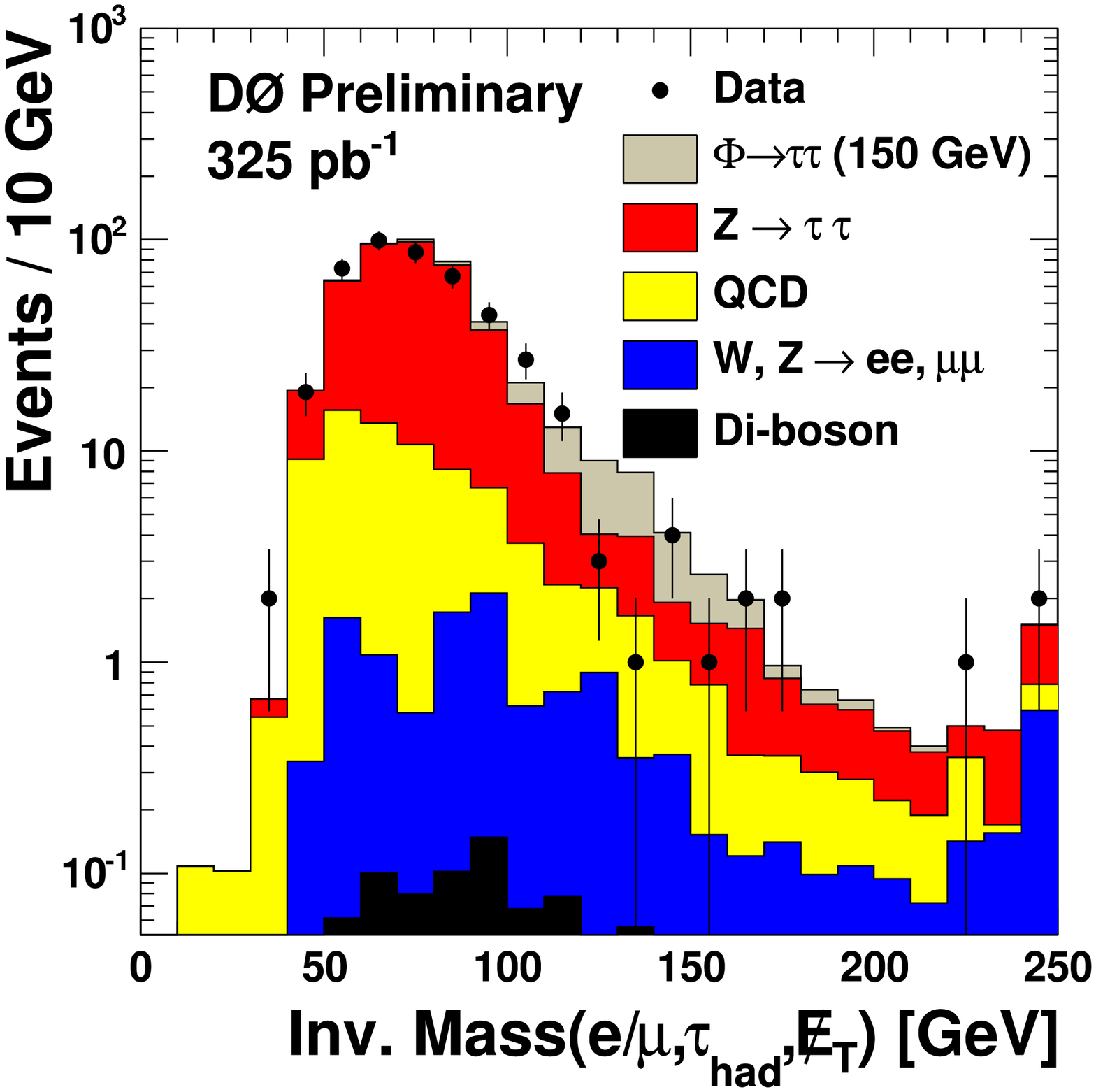}
&\includegraphics[width=1.75in]{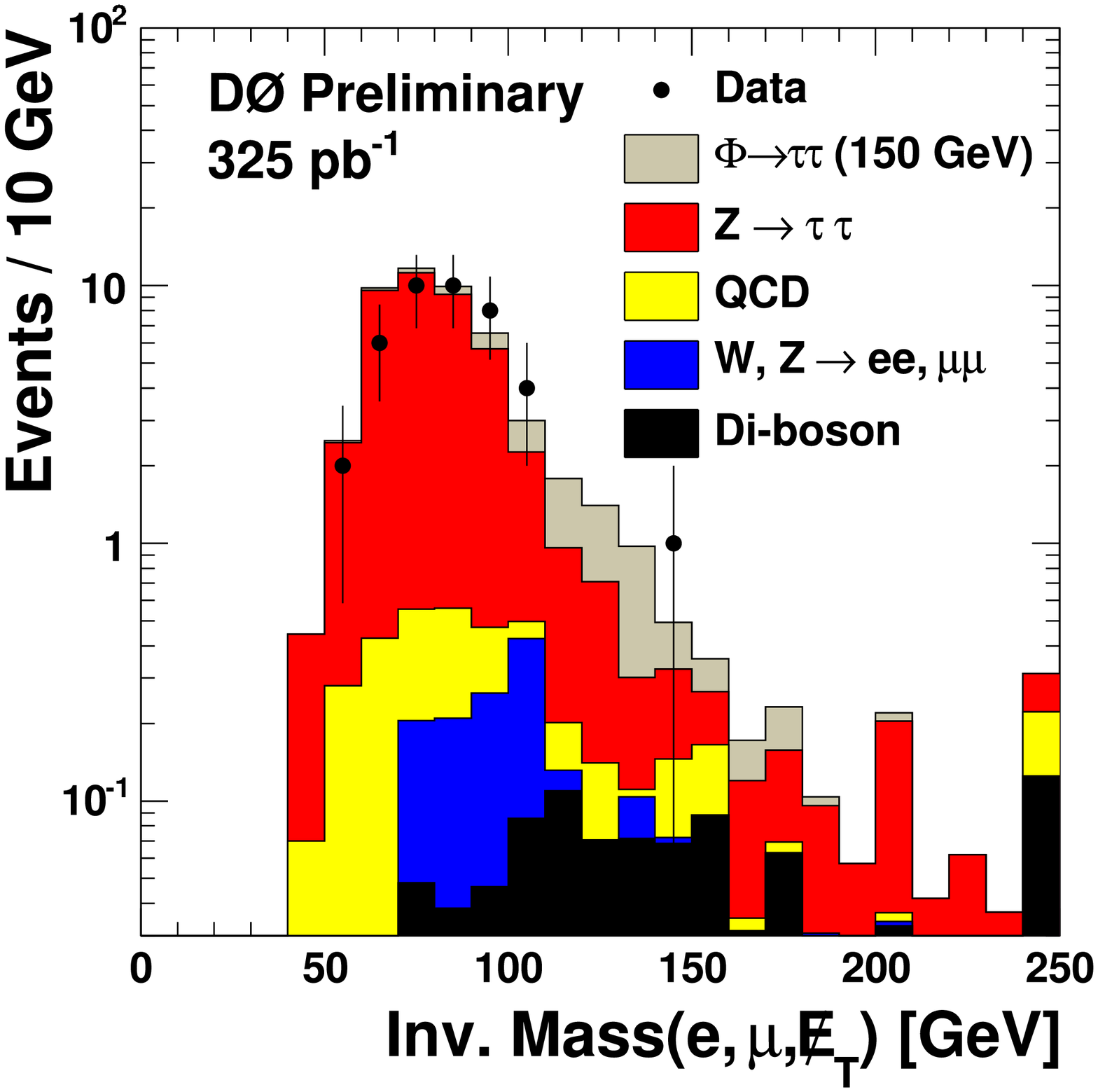}
&\includegraphics[width=1.75in]{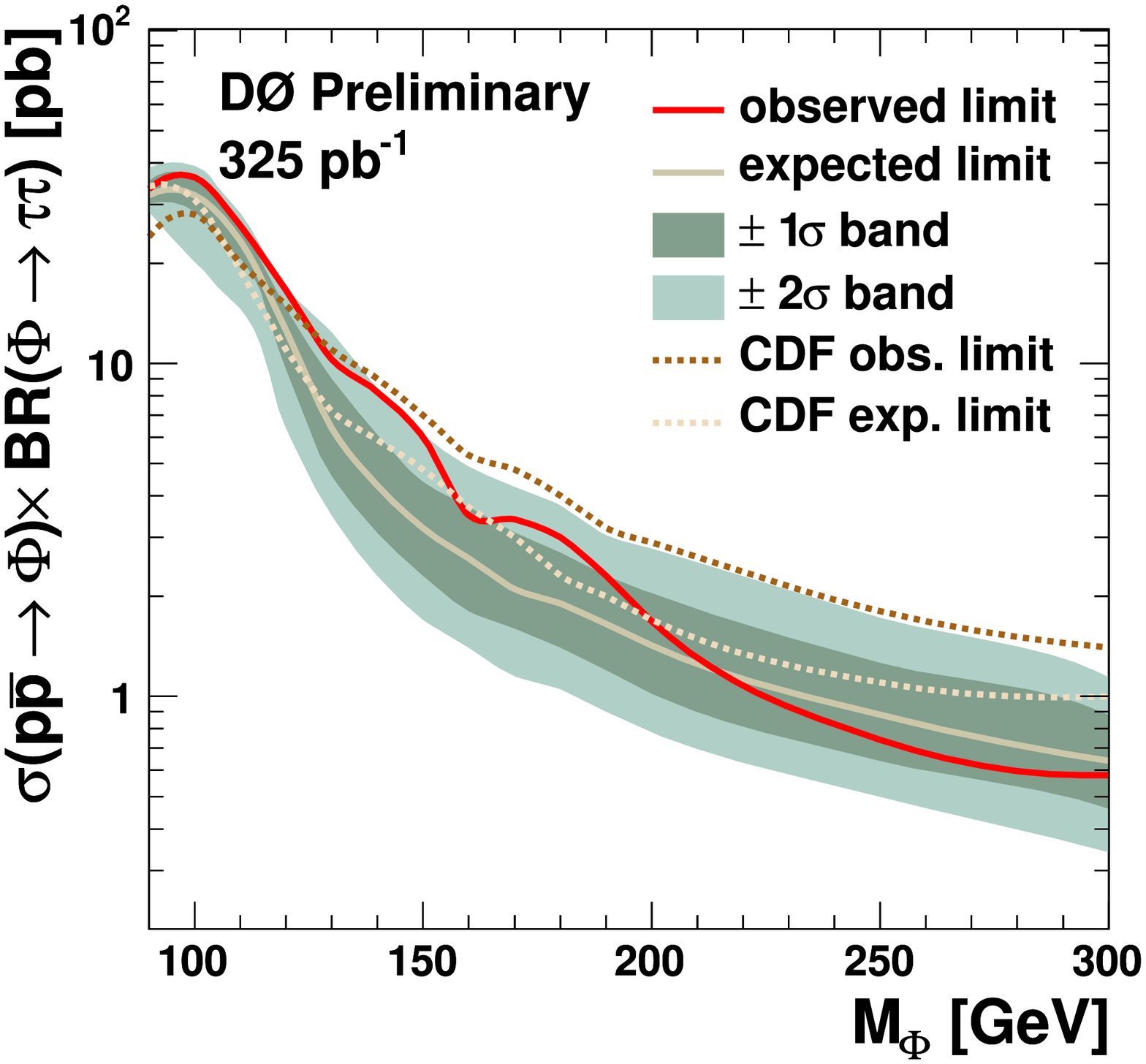}\\
(a) &(b)& (c)\\
\end{tabular}
\label{fig:htautauxsec}
\caption{The distribution of reconstructed visible mass
$\rm M_{vis}$ (see text) for (a) $ e\tau_h$+$ \mu\tau_h$
channel, and (b) $ e\mu$ channel. (c) The upper limits on
the production cross section as a function of Higgs mass at
95\% CL.}
\end{figure}

The reconstructed visible mass defined as
$\rm M_{vis}=\sqrt{(P_{\tau 1}+P_{\tau 2} +\mpt)^2}$,
where $\rm P_{\tau 1,2}$ is the four vector of the
visible tau decay products and $\rm \mpt=(\met ,\metx , \mety ,0) $,
is used to discriminate between signal and background processes.
No excess over the background expectation has been
observed.
The Figs.~\ref{fig:htautauxsec}(a)-(b) show the
distributions of $\rm M_{vis}$ (denoted as 
``$\rm Inv.\  Mass(e/\mu,tau_{had},\met)$" and
``$\rm Inv.\  Mass(e, \mu,\met)$" respectively in the figure)
for different final states,
while Fig.~\ref{fig:htautauxsec}(c) displays the upper limits
on production cross section as a function of Higgs mass.
\section{Combined Limits}
\begin{figure}[h]
\begin{tabular}{cc}
\includegraphics[height=.2\textheight]{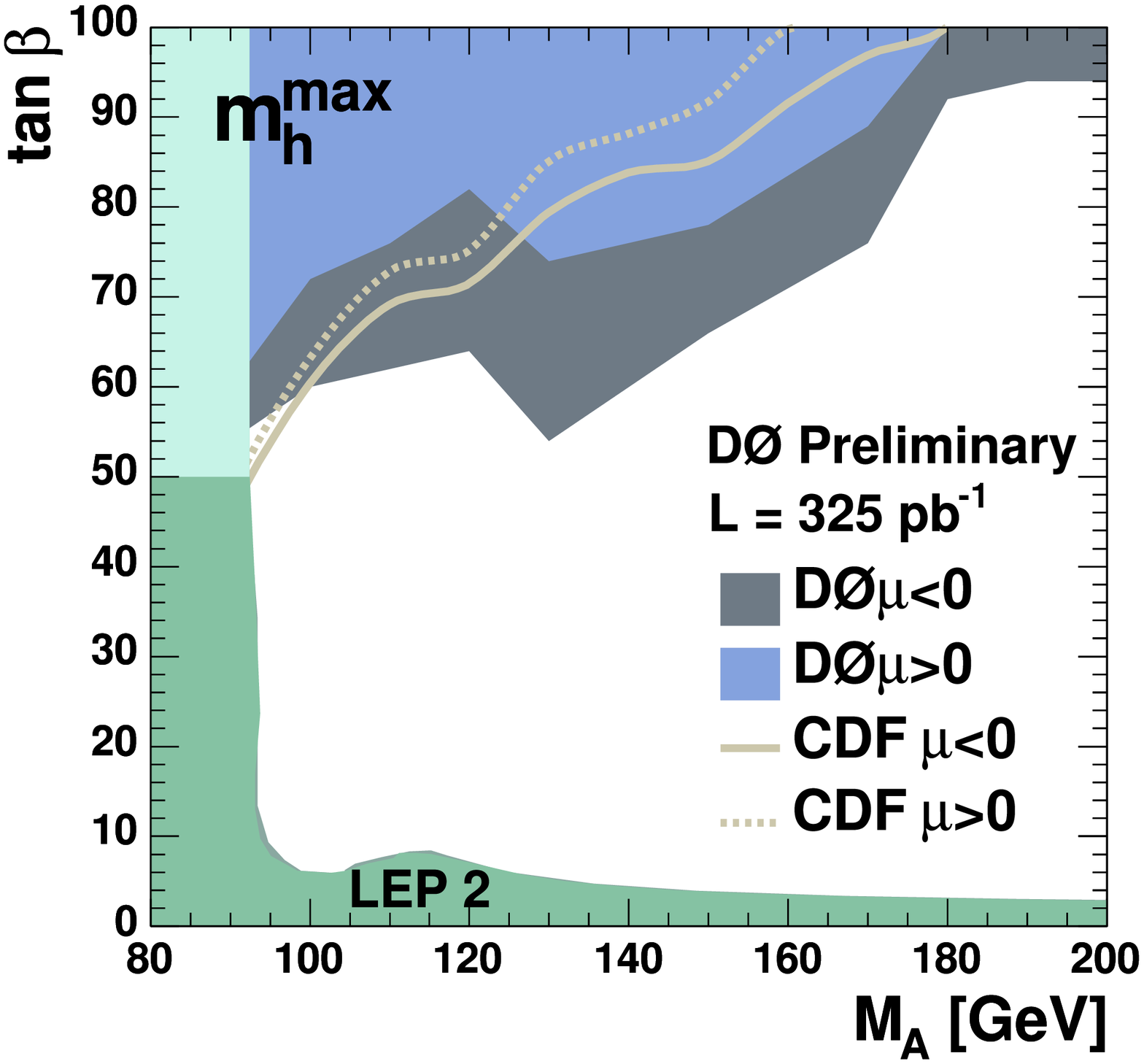}
&\includegraphics[height=.2\textheight]{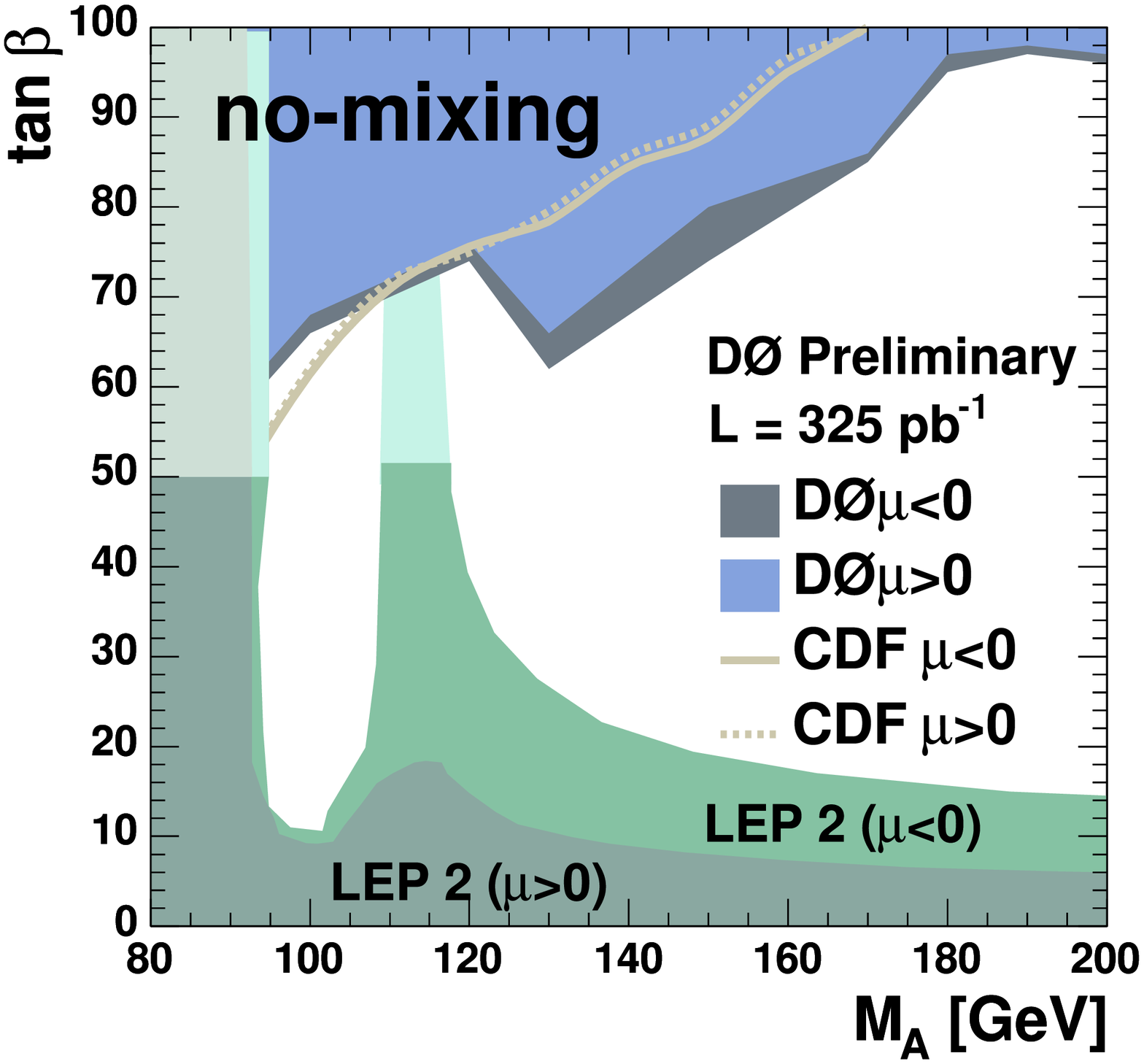}\\
(a) &(b)\\
\end{tabular}
\label{fig:combinedlimit}
\caption{The excluded region in the ($\rm M_A, tan\beta$)
plane for the $\rm m_h^{max}$ and no-mixing scenario
with $\rm \mu=\pm 200$ GeV.}
\end{figure}

The D\O\ results obtained from the above two analyses
\cite{prlhbb}\cite{prlhtt}
are combined together and are re-interpreted in
``$\rm m_h^{max}$'' and ``no-mixing'' scenarios.
Fig.~\ref{fig:combinedlimit} shows the current D\O\  limits
on MSSM parameter space. The similar limits obtained
by the CDF~\cite{cdfhtautau} are also displayed for the purpose
of comparison.
\section{Conclusions}
D\O\  has performed searches for the MSSM neutral
Higgs bosons and the current D\O\  limit on MSSM parameter
space is the most sensitive to date. Additional 
search channels like
$p\bar{p}\rightarrow\phi(\rightarrow \tau^+\tau^-)b(\bar{b})X$,
advanced analysis techniques and larger dataset will 
provide further scope to improve the sensitivity in
future.

\end{document}